\documentclass[a4paper]{spicawStyle}
\usepackage{graphicx,natbib}

\newcommand{\OI}{O\,{\sc i}}
\newcommand{\HI}{H\,{\sc i}}
\newcommand{\OIII}{O\,{\sc iii}}

\newcommand{\NIII}{N\,{\sc iii}}
\newcommand{\CII}{C\,{\sc ii}}
\newcommand{\SiII}{Si\,{\sc ii}}
\newcommand{\FeII}{Fe\,{\sc ii}}
\newcommand{\HII}{H\,{\sc ii}}
\newcommand{\NeIII}{Ne\,{\sc iii}}
\newcommand{\SIII}{S\,{\sc iii}}
\newcommand{\SI}{S\,{\sc i}}
\newcommand{\FeIII}{Fe\,{\sc iii}}
\newcommand{\ArIII}{Ar\,{\sc iii}}
\newcommand{\PIII}{P\,{\sc iii}}
\newcommand{\NeII}{Ne\,{\sc ii}}
\newcommand{\SIV}{S\,{\sc iv}}
\newcommand{\ArII}{Ar\,{\sc ii}}
\newcommand{\NiII}{Ni\,{\sc ii}}

\begin{document}

\title{The interstellar gas seen in the mid- and far-infrared:\hspace{4cm}
The promise of SPICA Space Telescope}

\author{Javier R. Goicoechea\inst{1} and Jos\'e Cernicharo\inst{1,2}} 

\institute{Centro de Astrobiolog\'{\i}a, CSIC-INTA, Madrid, Spain
\and \textit{Herschel Space Observatory} Mission Scientist}

\maketitle 

\begin{abstract}

The mid- and far-IR spectral ranges are critical windows to characterize the physical 
and chemical processes that transform the interstellar gas and dust into stars and planets. 
Sources in the earliest phases of star formation  and 
in the latest stages of stellar evolution  
release  most of their energy 
at these wavelengths. Besides, the mid- and far-IR ranges provide key spectral diagnostics 
of the gas chemistry (water, light hydrides, organic species ...), of the prevailing physical 
conditions (H$_2$, atomic fine structure lines~...), and of the dust mineral and ice 
composition that can not be observed from ground-based telescopes. 
With the launch of JAXA's SPICA telescope, uninterrupted studies
 in the mid- and far-IR 
will be possible since  ESA's \textit{Infrared Space Observatory} (1995). In particular,
SAFARI will provide full access to the $\sim$34-210\,$\mu$m waveband through
several detector arrays and flexible 
observing modes (from broadband photometry
to medium resolution spectroscopy  with $R$$\sim$3,000 at 63\,$\mu$m), and
reaching very high line 
sensitivities ($\sim$10$^{-19}$ \,W\,m$^{-2}$, 5$\sigma$-1hr)
within a large FOV ($\sim$2$'$$\times$2$'$). 
Compared to previous far-IR  instruments (ISO /LWS,  \textit{Akari}/FIS, 
\textit{Spitzer}/MIPS and \textit{Herschel}/PACS), SAFARI will provide a superior way to obtain 
fully-sampled spectro-images 
and continuous SEDs of very faint and extended ISM sources 
 in a wavelength domain not accessible to JWST or ALMA.
The much increased sensitivity of SPICA will allow us to step forward and 
reveal not only the chemical complexity  in the local ISM, but also in the extragalactic ISM
routinely.

\end{abstract}

\section{Introduction}

Due to the atmospheric opacity, the mid- and far-infrared  domains have been the 
latest spectral windows used in astrochemistry. The potential of opening a 
new spectral wavelength range for molecular spectroscopy begun to be fully exploited 
by  ESA's \textit{Infrared Space Observatory}, ISO (Kessler et al. 1996), 
it has been successfully  continued  by NASA's \textit{Spitzer} (Werner et al. 2004) and 
JAXA's \textit{Akari} (Murakami et al. 2007),
 and it will further advance with ESA's \textit{Herschel Space
Observatory} (Pilbratt et al. 2009). 
Before ISO, the mid- and far-IR spectra of the most relevant 
interstellar (ISM) and circumstellar (CSM) sources were  unknown. 
In particular, most gas phase species emitting in the mid- and far-IR were unidentified. 
Nowdays we know that the spectral features that can be detected in 
this  domain are essential to study  the physical conditions and the
chemical content of a broad range of astronomical environments. Such diagnostics include 
atomic fine structure lines, PAH bands, H$_2$ lines, light hydrides and many 
non-polar molecules 
that can not be detected with ground-based radio telescopes 
(\textit{e.g.,} PdBI, SMA or ALMA and SKA in the future). 
Besides, the mid- and far-IR provide access
to a great variety of ice and mineral features that help us to understand the 
life-cycle of cosmic dust.

In this short contribution we argue why a very sensitive space telescope covering
the full mid- and far-IR range can decisively contribute to improve our understanding of the
chemical complexity in the Universe. We also review some representative examples of 
previous mid- and far-IR observations
towards very bright ISM and CSM sources in the Milky Way. 
These  spectra remain unique examples of the diagnostic power of this 
domain, and are  archetypes of what much more sensitive space telescopes 
(SPICA!) will routinely observe in much fainter regions of the sky,
and that are beyond
the sensitivity capabilities of \textit{Herschel}.
The contribution of dust grains and PAHs to the mid- and far-IR spectrum of different
environments is well covered 
by other papers in these proceedings. Here we shall put more emphasis on gas phase molecular 
spectroscopy. More  detailed reviews can be found in \textit{e.g.,} van~Dishoeck~(2004)
or  Cernicharo \& Crovisier~(2005).

\begin{figure*}[ht]
  \begin{center}
    \includegraphics[width=12 cm]{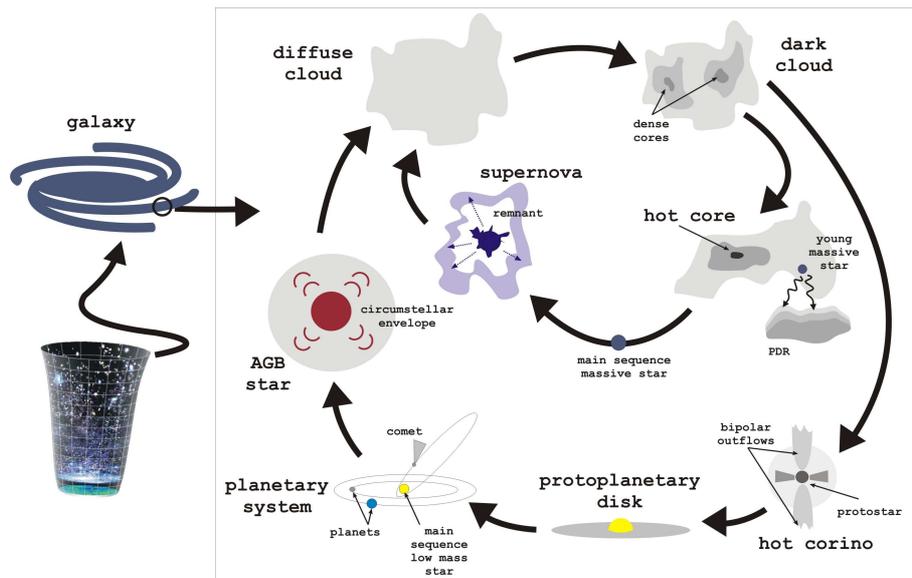}
  \end{center}
  \caption{Gas and dust life-cycle in the Molecular Universe. The life-cycle of interstellar
  matter is tied to stellar evolution and the formation of planetary systems. 
  SPICA will follow the gas, mineral and ice
  inventory, as well as their formation, processing and destruction 
  mechanisms by targeting all relevant 
  environments  in the Milky Way and in external galaxies. From Ag\'undez (2009).}
\label{goicoecheaj1_fig:life-cycle}
\end{figure*}

\section{Why another mid- and far-IR space telescope?}
\label{goicoecheaj1_sec:tit}

The life--cycle of interstellar gas and dust has been traditionally studied through line absorption 
observations  in the UV and visible domains (\textit{e.g.,} the diffuse and translucent ISM)
and through
submm, millimeter and centimeter 
 observations with ground-based radiotelescopes (usually by detecting emission lines). 
Due to the atmospheric opacity  and because of the nature 
of IR detectors,  observations over the entire mid- and far-IR bands require space telescopes 
and cryogenic 
instrumentation (down to a few mK!). Both requirements imply
a significant technological (and financial) 
challenge, as well as a notable international collaboration. All of them
can be better justified  
if:\\ 
\textbf{1)} Science drivers are exciting and unique 
(hard to achieve with other instrumentation). 
\textbf{2)} Instruments provide outstanding scientific results 
(with great impact in the field).\\

In (gas phase) astrochemistry there are three particular areas  that are \textbf{only} accessible
in the mid- and far-IR:

\textbf{Fine structure lines} from key gas reservoirs
 such as [\FeII]26, [\SiII]35, [\OI]63,145 and [\CII]158$\mu$m 
($E_i<$13.6\,eV) that are critical diagnostics of the neutral gas component of the ISM 
and are often the most important  cooling lines. 
On the other hand, fine structure lines of ions with $E_i>$13.6\,eV
([\OIII]52, 88, [\NIII]57\,$\mu$m, ...) trace the ionized gas component 
(\textit{e.g.,} \HII\, regions:
electron densities, ionizing radiation strength, etc.).
All these lines are usually bright and are not affected by extinction 
problems  as they are at shorter wavelengths (visible and near--IR).

\textbf{Light molecules} 
that have their pure rotational transitions in 
this range: \textit{e.g.,} H$_2$, water vapor, OH, HD, H$_2$D$^+$ and many other hydrides. 
Together with the above atomic fine structure lines, their spectral signature
allows one to penetrate  heavily obscured
regions like star forming regions or the ISM in galactic nuclei, and provide means
to derive the 
physical conditions (densities, temperatures, UV fields) and the chemical composition in 
harsh environments (\textit{e.g.,} UV irradiated regions, XDRs,
shocked-regions, etc.). Heavier species (CO, HCN, ...) have their high energy transitions in 
this domain (their ``high-$J$ lines") and  trace even higher temperature and density phenomena
 (\textit{e.g.,} strongly shocked-regions such as SNe remnants, 
the inner circumstellar shells around evolved stars, etc.). 

\textbf{Organic chemistry}: symmetrical molecules (\textit{i.e.,} all species
lacking a permanent electric dipole)
do not have pure rotational lines (\textit{i.e.,} to be observed from  radiotelescopes)
but do show vibration-rotation transitions in the mid- and far-IR 
(stretching and bending modes). Very relevant and abundant species of this kind are, 
for example, acetylene (C$_2$H$_2$), methane (CH$_4$), carbon dioxide (CO$_2$),
 benzene (C$_6$H$_6$), many different pure 
carbon chains (C$_n$) and of course, the ubiquitous 
polycyclic aromatic hydrocarbons (PAHs).

\section{The interstellar gas and dust  life-cycle}
 The first of the four Cosmic Vision (CV) Themes proposed by ESA, 
\textit{``What are the conditions for planet 
formation and the emergence of life?"} calls for a mission able to 
\textit{``investigate star-formation areas 
and protostars..."} and  \textit{``investigate the conditions for star formation and evolution"}. 
As stated  in the CV objectives, \textit{``we still lack a comprehensive theory explaining why and 
how stars form from interstellar matter and, apparently quite often, planetary systems with them"}. 
SPICA instruments will make decisive contributions to the study of the physical and chemical 
processes that transform the \textit{``gas and dust into stars and planets"}, or in other words, 
it will follow the life-cycle of matter: from the formation of 
dust grains in circumstellar shells,
  to its incorporation to planet forming 
disks, some of them maybe similar to what our own Solar System looked like
at the time of its formation (see the gas and dust life-cycle diagram in
Figure~\ref{goicoecheaj1_fig:life-cycle}). 

SPICA observations will allow us to significantly improve our knowledge of many
areas related with planetary formation and evolution 
(see detailed contributions in these proceedings). 
As a unique complement, SPICA will decisively contribute
to unveil the initial conditions for star formation 
(a mission able to \textit{``map the birth of stars and planets by peering into the highly 
obscured cocoons where they form"}) and in the processes associated with stellar death 
(evolved stars, supernova remnants, etc.) that both enrich and process the ISM gas and dust 
for a  subsequent generation of star formation.

The process of molecular cloud formation, star formation and star death are poorly understood 
because of the large amounts of dust obscuring the molecular cloud sites where stars are born,  
as well as the circumstellar envelopes (CSE) of molecular gas and dust created at the end of  
Sun-like stars' life. Since the dust becomes ``transparent" at far-IR wavelengths,  
this spectral window provides the natural way to penetrate inside 
these environments.
By observing in the mid- and far-IR, SPICA will provide and unprecedented window into key aspects 
of the gas and dust life-cycle: from its formation in evolved stars, its evolution in the ISM, 
its processing in supernova-generated shock waves and near massive stars (winds, \HII~regions, etc.) 
to its final incorporation into star forming cores, protostars, protoplanetary disks and 
eventually planets.

   Very important gas-phase species that can not be observed from the ground 
have their spectral signatures in the operative  range of SPICA.
Such spectral diagnostics provide
clues on the elemental  abundances of key elements  (C, O, D, Si, ...) and they also provide deep 
insights into the  gas/dust chemical interplay: deuterium enrichment, ice 
formation, grain growth, evaporation, photodesorption or metal depletion on 
grains (\textit{e.g.,}~Okada et al. 2008).

\section{Local and extragalactic ISM with SPICA}

 An important fraction of the interactions that transform the gas and dust 
into dense molecular clouds and stars, takes place over very large 
scale regions, several arc minutes size 
(\textit{e.g.} diffuse ISM clouds, cirrus, high latitude clouds, star forming regions, 
circumstellar shells, 
supernova  remnants, etc.) that are too faint to be mapped entirely by \textit{Herschel} in 
spectroscopy mode, 
or too extended to be mapped with ground-based interferometers.  
In this context, the niche of SPICA/SAFARI is to provide far-IR spectro-images 
(of many key spectral diagnostics simultaneously) of the regions that are too obscured for 
JWST to examine or too warm/extended to be efficiently traced with ALMA in the millimeter  
domain. 

 SPICA instruments have the appropriate combination of spatial and spectral resolution 
together with high sensitivity and large field-of-view (FOV) to follow the evolution of 
interstellar matter over very large spatial scales  (\textit{e.g.,}~by taking full SEDs 
and spectra of distant star forming regions), study meaningful samples of compact sources
(\textit{e.g.,}~hundreds of protostars, distant evolved stars and disks around 
all types of stars and ages) and out to large distances (inaccessible to
previous telescopes).

In the following we just list several fields where we 
anticipate SAFARI can play a unique role 
(difficult to achieve with other telescopes):

\begin{itemize}
\item Spectro-imaginery of the ISM in nearby galaxies.

\item Chemical complexity beyond our galaxy.

\item Continuous SEDs of Young Stellar Objects. 

\item  High-mass star formation in distant SF regions.

\item Evolved stars and interaction with ISM.

\item Supernova remnants (dust formation and processing?).

\item Molecular cloud formation.

\item The first identification of PAH molecules in space.
\end{itemize}

Thanks to its much improved sensitivity, SPICA will extend our knowledge of the
physics and chemistry of the ISM matter in our galaxy to similar
detailed studies of the ISM in nearby, spatially resolved galaxies
(see \textit{e.g.,} Gonz\'alez-Alfonso et al. 2004; Goicoechea et al. 2005).

\begin{figure*}[ht]
  \begin{center}
    \includegraphics[width=11.4 cm, angle=-90]{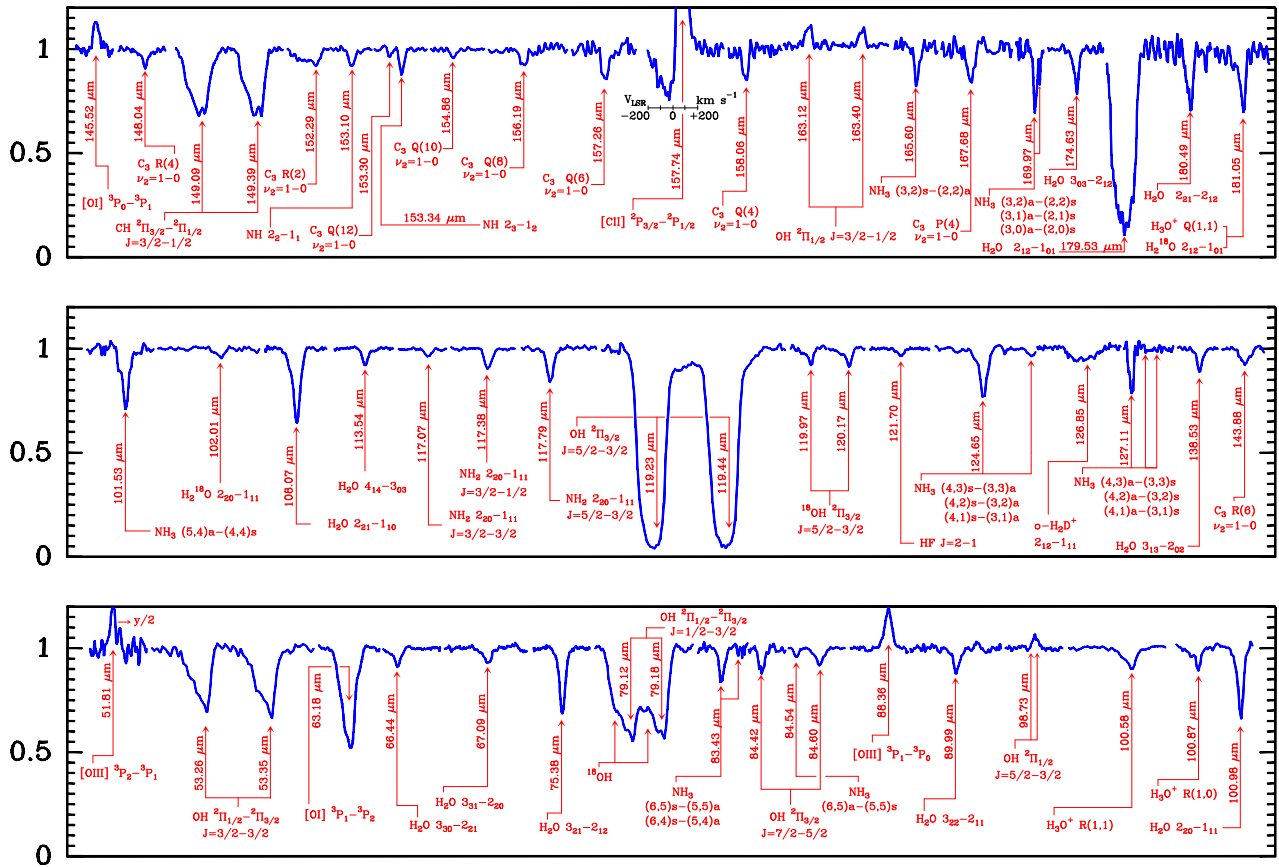}
  \end{center}
  \caption{Far-IR spectrum of Sgr B2 star forming region
  in the Galactic Center taken with the ISO/LWS.
  Line detections have been discussed in detail in different 
  works (see Goicoechea et al. 2004; Polehampton et al. 2007 and references therein).}
\label{goicoecheaj1_fig:sgrb2m_iso}
\end{figure*}

\subsection{What will not be done by Herschel \& JWST?}

The successful launch of ESA's \textit{Herschel Space Observatory} in May~2009 
-- the biggest IR telescope ever launched~-- suggests that \textit{Herschel} will be 
a major breakthrough for far-IR and submm astronomy. At the time of writing, 
commissioning has been
declared successful and the instruments are now in the performance verification phase.
However, \textit{Herschel} has several constraints and there will be many critical areas
of the ISM and CSM research  that will not be covered due to limited sensitivity, 
wavelength coverage or spectral resolution:

\begin{itemize}
\item \textit{Herschel} observes the $\sim$55 to 672\,$\mu$m waveband so it does not access the
  mid-IR domain where most mineralogy and ice studies can be carried out. Similarly, it can not
  observe the mid-IR emission of PAHs and most organic molecules or the high excitation 
  transitions (high-$J$) of many species. SPICA, however, will cover the full 
$\sim$5 to 210\,$\mu$m  waveband (baseline).
\item \textit{Herschel} is a ``warm'' telescope which greatly limits
  its sensitivity and thus mapping speed to obtain spectro-images of very
  extended/distant ISM sources. SAFARI, the far-IR imaging-spectrometer 
  for SPICA, will cover the critical $\sim$34-210\,$\mu$m that is not accessible to JWST or ALMA with 
  a factor $\sim$200 (in photometry) and $\sim$20 (in spectroscopy) better sensitivity than
   \textit{Herschel} /PACS, and with a wider instantaneous FOV
  (\textit{i.e.,} $\sim$2$'$$\times$2$'$, a factor $\sim$6.5 larger).

\item  JWST/MIRI ($>$2014) will operate in the mid-IR ($\sim$5-28\,$\mu$m band), 
so it does not cover the far-IR. MIRI will provide imaging
 and medium resolution spectroscopy ($R$$\sim$3,000) over a small FOV (a few arcsec$^2$). 
SPICA's mid-IR cameras and spectrometers will provide
larger FOVs and out to an order of magnitude higher spectral resolution ($R$$\sim$30,000)
in selected mid-IR wavebands. 

\end{itemize}

Despite its modest monolithic mirror size ($D$$\sim$3.5m,  
similar to \textit{Herschel}) the SPICA telescope will be actively cooled down to
$\sim$4.5 K (compared to $\sim$45 K through passive cooling for JWST), 
and thus will reach similar sensitivities than
JWST/MIRI in the mid-IR (around $\sim$1\,$\mu$Jy) but with a much increased wavelength 
range (including the far-IR!), wider FOV and increased
observing efficiency due to the use of multiple detector arrays.

\subsection{What will not be done by ALMA?}

The \textit{Atacama Large Millimeter Array} (ALMA) will be fully operational the next decade.
With its extremely high angular resolution and huge collecting area, ALMA will be a major
revolution
in many fields of the cold and molecular Universe. However, ALMA observes at wavelengths
longer than $\sim$315\,$\mu$m ($<$950\,GHz) and for practical reasons (it is a very long-baseline interferometer)
it is not designed to carry out extensive surveys of very large regions of the sky. In addition,
there are critical spectral diagnostics (at their restframe wavelengths) 
that are not accessible to ALMA:

\begin{itemize}

\item The study of the oxygen chemistry will be limited to observations of CO, HCO$^+$ and
 other trace species, but it will not observe the  emission of the
 major oxygen reservoirs (\textit{e.g.,} atomic oxygen, water ice, CO$_2$ or the thermal
 line emission of water vapor).

\item The most important atomic and ionic fine structure lines are outside the coverage 
of ALMA, which will limit the determination
of the physical conditions (temperatures, densities and radiation fields)
and the study of the energy balance (heating and cooling) 
in PDRs, XDRs, and shock-like environments.

\item ALMA can only detect the mm/submm dust continuum emission but
it can not observe the specific grain and ice spectral features that are needed to determine
the dust composition and its formation history along the  life-cycle of interstellar matter.

\item ALMA will not even touch several aspects of the chemical complexity in the Universe.
 In particular, it will not observe molecules without permanent electric dipole
 or the mid-IR emission of PAHs, thus missing a significant fraction of 
the organic chemistry in Space.

\end{itemize}

Note, however, that the above  IR spectral diagnostics  become 
 available to ALMA in highly-redshifted extragalactic sources. In particular, ALMA will
detect the restframe mid- and far-IR spectrum of sources at redshifts of 
$z>\frac{315}{\lambda_{rest}\,(\mu m)}-1$. For example, $z>4$ for the [\OI]63\,$\mu$m line
and $z>10$ for the H$_2$ 0-0 $S$(0) line at $\lambda_{rest}$=28\,$\mu$m.

\section{Examples of mid-IR and far-IR spectroscopy}

We now review some interesting results of previous mid- and far-IR observations
towards  Sgr B2 and Orion~KL high-mass star forming regions (SFRs).
Sgr~B2 ($\simeq$8.5\,kpc) and Orion KL ($\simeq$450\,pc) can be considered
as two of the most remarkable giant molecular clouds for astrochemistry  studies. 
They are also very appropriate sources to 
construct a  \textit{template} for more distant (\textit{e.g.,} fainter) and 
unresolved regions (\textit{e.g.,} extragalactic).
The cores of both SFRs are the most  prolific sites of molecular line 
emission/absorption in the Galaxy (in terms of density and intensity/absorption depth of lines). 
Due to the large column density of warm molecular material  (Sgr~B2) 
or due to its proximity (Orion), both complexes are among the brightest far--IR 
sources of the sky. For these reasons, Sgr~B2 and Orion~KL were
fully surveyed between $\sim$2.4 and 197\,$\mu$m by ISO's SWS and LWS
 spectrometers.
The resulting spectra remain unique examples of the diagnostic power of this domain,
and are archetypes of what SPICA  could routinely observe in much fainter
sources (protoplanetary disks, outer Solar System bodies, etc.).

\subsection{The mid- and far-IR spectrum of a SFR}

The far--IR spectrum (47-196\,$\mu$m) of Sgr~B2
(see Figure~\ref{goicoecheaj1_fig:sgrb2m_iso}; Goicoechea et al.~2004; Polehampton et al.~2007
and references therein) contains:
$(i)$ The peak of the thermal emission from dust
(a blackbody at $\sim$30\,K roughly peaks at $\sim$100\,$\mu$m); 
$(ii)$~Atomic fine structure lines that are 
major coolants of the warm gas, and excellent discriminants of 
the PDR, H\,{\sc ii} and shock emission
([O\,{\sc i}], [O\,{\sc iii}], [C\,{\sc ii}], [N\,{\sc ii}], [N\,{\sc iii}]);
$(iii)$ Rotational lines from key  light hydrides
(H$_2$O, OH, H$_3$O$^+$, CH$_2$, CH, NH$_3$, NH$_2$, NH, HF, HD, H$_2$D$^+$) and
ro-vibrational lines from abundant nonpolar molecules (C$_3$). 
These species provide unique information on the prevailing physical conditions
and on the basic O, C, N and D chemistry.
Most of the molecular lines appear in \textit{absorption} whereas
atomic and ionic lines appear in emission (except for absorption in the 
[O\,{\sc i}]63 and [C\,{\sc ii}]158\,$\mu$m lines).
In particular, [O\,{\sc i}]63 and [C\,{\sc ii}]158\,$\mu$m absorption line
profiles provide clues regarding several peculiarities
observed in some extragalactic spectra at lower resolution
(\textit{e.g.,} the [C\,{\sc ii}]/far-IR deficit).
The main gas component traced by far--IR observations is the
warm, low density envelope of Sgr~B2 (T$_k$\,$\simeq$300\,K; 
$n$(H$_2$)$<$10$^4$\,cm$^{-3}$). 
Given the low densities in this component, no high--$J$ CO line was detected
at ISO's sensitivity.
This situation may apply to other \textit{warm} regions observed in the far--IR.
The warm, low--density gas is particularly difficult to
trace in the millimeter domain where one usually observes molecular emission lines
from collisionally excited gas. 
Finally, because of its location in the Galactic Center,
all ground--state lines towards Sgr~B2 show a broad absorption profile 
($\Delta v$\,$\simeq$200\,km\,s$^{-1}$) due to 
foreground absorption produced by the spiral arm diffuse clouds in the line of sight.

On the other hand, the far--IR spectrum of Orion~KL 
(Lerate et al.~2006 and references therein) is
dominated by \textit{emission} lines from molecular
(H$_2$O, high--$J$ CO, OH, NH$_3$) and atomic species.
Interestingly enough, H$_2$O and OH line profiles show a complex behavior
 evolving
from pure absorption, P~Cygni type, to pure emission, depending
on the transition wavelength, $E_{up}$ and line opacity. 
These lines arise from Orion outflow(s) and associated shocked regions. 
Without resolving these profiles, low resolution spectra of similar regions
may lead to a misinterpretation of the prevailing dynamics and physical conditions.
Given the high densities and temperatures of Orion's shocked gas,
CO pure rotational emission up to $J$=39 has been detected 
(with an upper energy level of E$_{up}/k$\,$\simeq$4,000\,K!!).

\begin{figure}[h]
  \begin{center}
    \includegraphics[width=6cm, angle=-90]{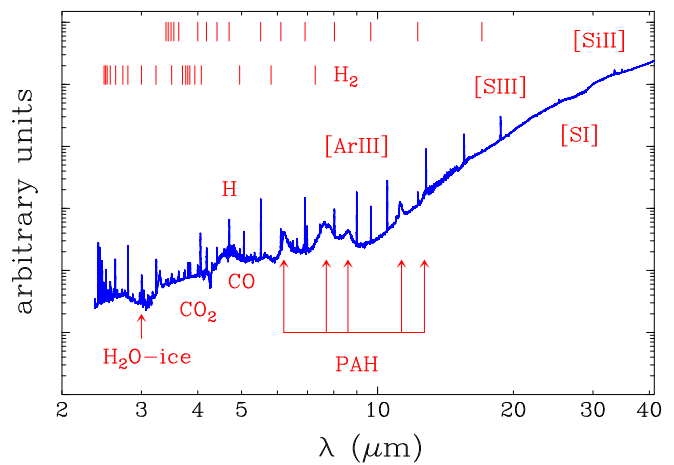}
  \end{center}
  \caption{Mid-infrared spectrum of Orion Peak 1 shock obtained in the ISO/SWS grating scan
 observing mode (R$\sim$2,000). Some of the strongest lines and bands 
are labelled. The continuum levels of the individual bands, 
which differ due to aperture changes, were adjusted to make the spectrum appear
continuous. Adapted from Rosenthal et al. (2000).}
\label{goicoecheaj1_fig:peak1sws}
\end{figure}

Fig.~\ref{goicoecheaj1_fig:peak1sws} shows the ISO/SWS (2.4-45\,$\mu$m) spectrum of 
the brightest mid--IR H$_2$  emission peak of the Orion outflow (see Rosenthal et al. 2000).
This is one of the most spectacular shocked regions in the galaxy. 
56 H$_2$ ro-vibrational (up to $E_u/k$ =14,000\,K) and pure rotational 
lines from the H$_2$ 0-0 $S$(1) to 0-0 $S$(25) lines are detected. 
The spectrum also shows a number 
of PAH bands, \HI~recombination lines, ro-vibrational lines of 
CO ($\nu$=1-0; $\sim$4.7$\mu$m) and H$_2$O ($\nu_2$=1-0; $\sim$6.5$\mu$m), and
many ionic fine structure lines 
([\NeIII]36, [\FeII]36, [\SiII]35, [\SIII]33, [\FeII]26, [\SI]25, [\FeIII]23, 
[\ArIII]22,[\SIII]19, [\FeII]18, [\PIII]18, [\NeIII]16, [\NeII]13, [\SIV]11,
[\ArIII]9, [\ArII]7 and [\NiII]7\,$\mu$m), most of them coming from 
the foreground \HII~region and its bounding PDR. 
Van Dishoeck et al. (1998) also observed the Orion-IRc2 position with the same spectrometer.
In addition, several broadband features associated to silicate grains and H$_2$O--ice and 
CO$_2$--ice
were  detected along the line of sight to the strong mid--IR continuum of IRc2.

\begin{figure}[ht]
  \begin{center}
    \includegraphics[width=8cm]{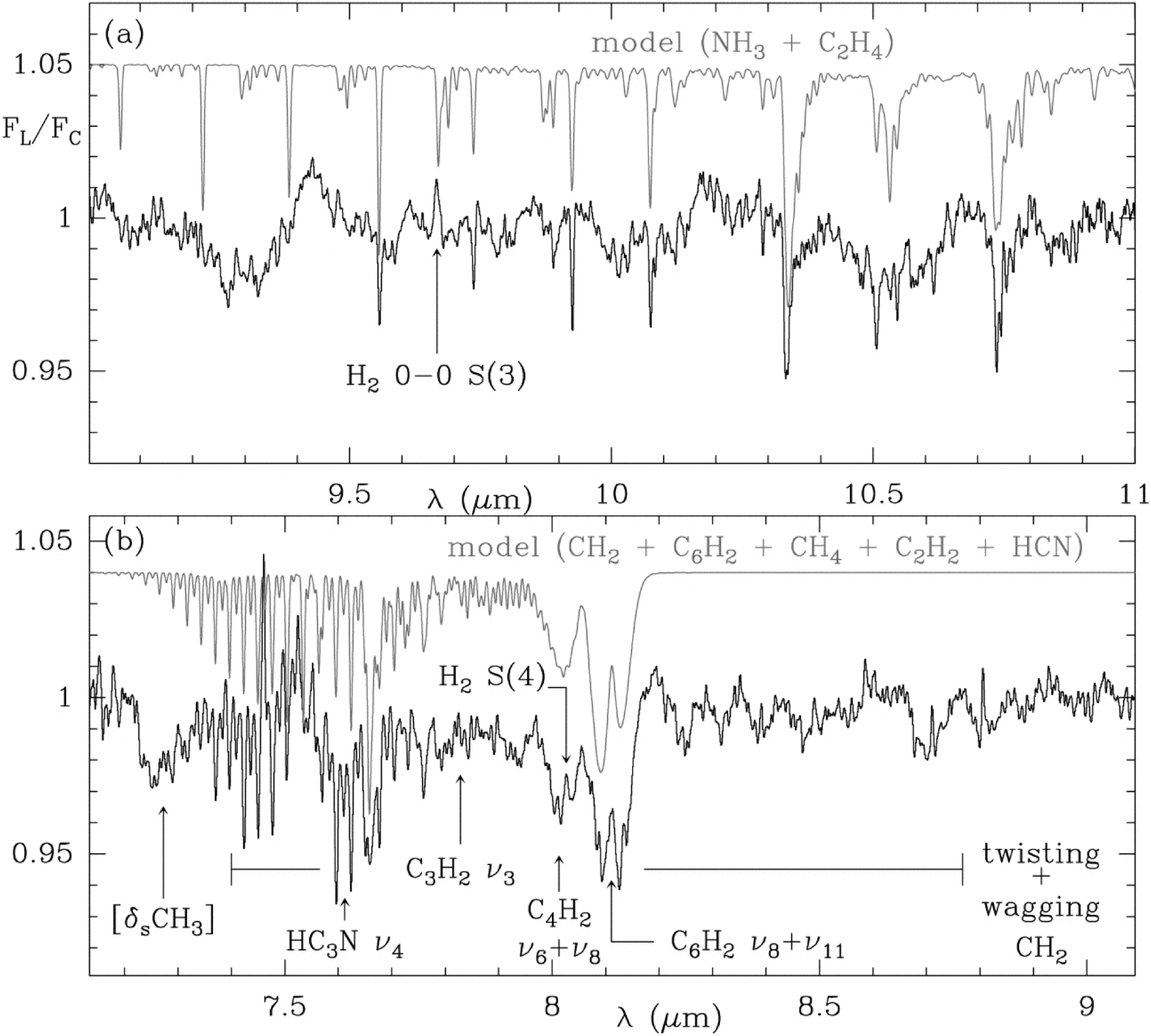}
  \end{center}
  \caption{Mid-IR spectrum of CRL~618 at selected wavelengths. 
The thin line in each panel corresponds to the model results discussed by 
Cernicharo et al. (2001a). 
The feature at 7.27$\mu$m corresponds to the symmetric bending mode of CH$_3$. 
The broad range of absorption bands produced by wagging and twisting modes of CH$_2$ 
is indicated by the horizontal line (Silverstein \& Webster~1998).
Figure from Cernicharo et al. (2001a).}
\label{goicoecheaj1_fig_carbons}
\end{figure}

\subsection{Mid-IR: Chemical complexity}

\begin{figure}[ht]
  \begin{center}
    \includegraphics[width=8cm]{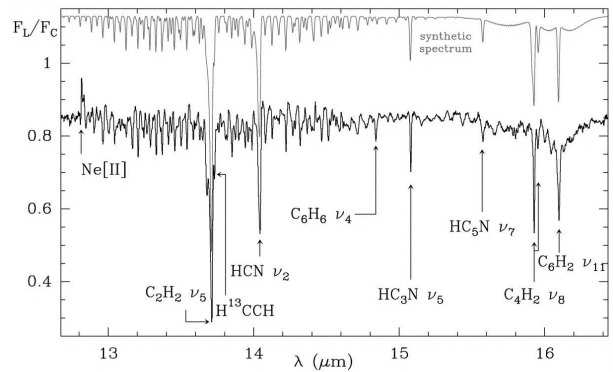}
  \end{center}
  \caption{Mid-IR spectrum of CRL~618 around 15$\mu$m. The infrared bands of 
the polyynes (C$_4$H$_2$, C$_6$H$_2$), benzene, and cyanopolyynes are indicated by arrows. 
The thin line corresponds to the synthetic spectrum  
 for C$_2$H$_2$, HCN, C$_4$H$_2$, C$_6$H$_2$, HC$_3$N, and HC$_5$N. 
All the narrow features to the right and left of the 
Q-branches of C$_2$H$_2$ and HCN correspond to the individual rovibrational lines of the 
P- and R-branches of these same two species. 
Figure from Cernicharo et al. (2001b).}
\label{goicoecheaj1_fig_benzene}
\end{figure}

Mid--IR spectroscopy (vibrational spectroscopy in particular) 
is a powerful tool to reveal the
chemical complexity in the Universe. A remarkable example is the organic composition
of  the circumstellar envelopes  around evolved stars.
Many organic species lack rotational spectra to be 
observed from radiotelescopes 
(\textit{e.g.,}~polyacetylenic chains; see Figure~\ref{goicoecheaj1_fig_carbons}). 
Benzene, the most basic aromatic unit,  
has been detected outside the Solar System only towards CRL 618 carbon-rich protoplanetary
nebula, PPNe  
(see Figure~\ref{goicoecheaj1_fig_benzene}; Cernicharo et al. 2001b) where the physical conditions 
in the inner CSE (shocks and UV photons) seem to drive the polymerization of acetylene and 
the formation of benzene.
This means that carbon-rich PPNes are probably the best organic chemistry factories in 
space and much can be learned about the gas/dust chemistry interplay 
(\textit{e.g.,} the formation of larger aromatic species or PAHs).  
SPICA's  very high-resolution mid-IR spectrometers will have the required sensitivity
to  study the organic content
and reveal the degree of chemical complexity, not only towards the brightest sources
in the sky,
but everywhere in the Milky Way and in nearby galaxies.

\section{Conclusions}

The mid- and far-IR domains provides key spectral diagnostics of the gas phase chemistry 
(\textit{e.g.,} water vapor, H$_2$, PAHs, organic species), of the prevailing physical 
conditions (atomic and ionic fine structure lines), and of the dust mineral/ice composition 
that can not be observed at other wavelengths.
Thanks to SPICA, uninterrupted studies in the critical $\sim$5-210\,$\mu$m range will be 
possible since 
the launch of ISO with orders of magnitude higher sensitivity and resolutions. 
Therefore, SPICA observations will revolutionize the study the life-cycle of interstellar matter in
nearby galaxies the same way ISO revolutionized the study of the ISM in the Milky Way in
the late 90's.

\begin{acknowledgements}
JRG thanks the European and Japanese SPICA science teams for fruitful discussions 
and different contributions to develop the ``gas and dust life-cycle" case.
We warmly thank Marcelino Ag\'undez for letting use his ``life-cycle figure".
JRG is supported by a \textit{Ram\'on y Cajal} research contract.

\end{acknowledgements}

\end{document}